\documentclass[12pt,preprint]{aastex}

\newcommand{\lsimeq}{{<\atop^{\sim}}}

\slugcomment{Version submitted to ApJL on  December 3 2002}
\shorttitle{{\em XMM-Newton} detection of WHIM at $z\sim 0.01$}
\shortauthors{Cagnoni}
\begin{document}
\title{{\em XMM-Newton} detection of warm/hot intergalactic medium  at  $z\sim 0.01$}
\author{I. Cagnoni\altaffilmark{1}}
\affil{Dipartimento di Scienze, Universit\`a dell'Insubria
Via Valleggio 11, I-22100, Como, Italy}

\begin{abstract}
We report a $\simeq 3 \sigma$ detection of a 21.82 \AA \/ absorption feature 
in the direction of the BL Lac MRK~421 which is interpreted as O~{\sc vii} 
K$\alpha$ at $z\simeq 0.01$. This corresponds to the redshift of a H absorber 
in a cosmic void detected by HST along the line of sight of the same object.
The 21.82 \AA  \/ line proves the existence of warm/hot intergalactic medium 
(WHIM) outside the Local Group, in agreement with current models 
of structure formation in the Universe.
The WHIM at $z\sim 0.01$ has temperature of $\lsimeq (2.5-4) \times 10^6$~K 
for gas densities in the range $10^{-6}-1$ atom cm$^{-3}$.
We also detect  O~{\sc vii},  O~{\sc viii} and  Ne~{\sc ix} absorption lines 
with zero velocity, possibly connected with WHIM in the Local Group.
\end{abstract}

\keywords{BL Lacertae objects: individual (MRK~421) --- intergalactic medium 
--- large-scale structure of universe --- quasars: absorption lines}

\section{Introduction}\label{sec_intro}
In the local Universe $\sim 50 -60$\% of the baryons visible at $z>2$ and 
predicted by the Standard Big-Bang nucleosynthesis are undetected \citep[e.g.][]{fuk98}.
Numerical simulations predict that such baryons, in the form of Ly$\alpha$ 
clouds (i.e. gas at $T <10^4$ K) and visible in the UV spectra of bright AGNs, 
have been shock-heated during the formation of the structures in the Universe to 
temperatures of $\sim 10^5-10^7$~K (i.e. the warm hot intergalactic medium, WHIM).
The most efficient way to detect such gas is through resonant absorption 
lines from highly ionized metals (e.g. O{\sc vi},O{\sc vii},O{\sc viii},Ne{\sc ix})
in the far UV and soft X-ray spectra of background sources \citep[e.g.][]{hel98,fan00}.
Current far UV observations proved the existence of the low temperature tail 
($T \sim (1-5) \times 10^5$ K) of the WHIM through the detection of O{\sc vi}
up to $z\sim 0.2$ \citep[e.g.][]{sem00, tri01}.
However the bulk of the WHIM baryons \citep[$\sim 70$\%, e.g.][]{fuk98}
are expected to lie at higher temperatures ($T\sim 10^6- 10^7$~K)
and their features should be detectable in the soft X-ray regime.
{\em Chandra} and {\em XMM-Newton} high resolution spectroscopy ($R \sim 100-500$)  
brought to the detection of 
a ``zero redshift'' O{\sc vii} absorption feature at 21.6 \AA \/
in the spectra of four bright AGNs  (PKS~2155-304, Nicastro et al. 2002; Fang et al. 2002; 
Cagnoni et al. 2003;  3C~273, Fang, Sembach \& Canizares 2003; 
H1821$+$643, Mathur, Weimberg \& Chen 2002 and MRK~421, Nicastro et al. 2001), 
interpreted as the signature of WHIM present within the Local Group.
The only published evidence of WHIM outside the Local Group is at $z\sim 0.05$ \citep{fan02}, 
however this feature is not confirmed in other {\em Chandra}  and {\em XMM-Newton} 
observations \citep[e.g.][]{nic02,cag03}.
In this paper I present  strong evidence for an  O{\sc vii} K$\alpha$ absorption line  
at  $z\sim 0.01$ in the spectrum of the BL Lac object MRK~421. \\
The paper is organized as follows:
\S~\ref{sec_obs} reports the {\em XMM-Newton} observations of MRK~421 and  describes the data reduction; 
\S~\ref{sec_whim} contains the spectral fits and the discussion on the observed WHIM absorption features. 
\S~\ref{sec_sum} is a conclusive section containing a summary of the results.

\section{Observations and  data reduction}\label{sec_obs}

MRK~421 is a calibration target for {\em XMM-Newton} and has been observed 
several times from the launch (Dec. 10, 1999) up to the time of writing.
Table~1 summarizes all the public  observations as of 
December 3, 2002.
MRK~421  X-ray spectrum is thought to be relativistically beamed synchrotron emission 
from energetic electrons and its intrinsic lack of features is the ideal laboratory 
to search for faint WHIM absorption lines.
In this paper I will concentrate on the high resolution ($\Delta E / E$ from 
100 to 500, FWHM, or 100 to 800, HEW, in the energy range 0.33--2.5 keV - 
5--38 \AA) Reflection Grating Spectrometers (RGS1 and RGS2) data collected 
by the two {\em XMM-Newton} X-ray telescopes\footnote{I refer the reader to 
\citet{sem02} for a detailed description of the observations and for and accurate spectral 
and timing analysis of EPIC data collected in 2000 and 2001.}.
{\em XMM-Newton} RGS effective areas  are complex in 
shape, and contain tens of narrow dips due to bad or hot columns or pixels in the CCD detectors, 
which require extremely accurate 
calibration measurements for a proper modeling \citep[e.g.][]{cag03}. 
Current RGS calibration uncertainties are as accurate as $\sim 5-10$\% between 7
and 36 \AA \/ and, as a consequence, false absorption/emission 
features with such relative intensities, are expected in the RGS spectra in physical units, 
in proximity of the known instrumental features. 
The strongest resonant absorption lines from neutral and/or 
highly ionized O and Ne, fall in wavelength ranges (i.e. 13-14 \AA, 18-20 
\AA, 20-24 \AA) in which RGS-2 spectra either do not exist (20-24 \AA, 
due to the failure of a CCD chip) or contain strong line-like shaped 
instrumental features \citep[see][]{cag03}. Therefore I rely on  RGS-1 
spectra, which, instead  are relatively 
instrumental-feature-free in these wavelength ranges, and use RGS-2
to double check the reality of a line, when possible, and to cover the 
 10.5-14.2 \AA \/ region, where RGS-1 has a failed CCD chip.
I also restrict our analysis to the first order spectra only.

I reprocessed the data using {\em XMM-Newton Science Analysis System} (SAS)
version 5.3.0 and the latest calibration files as of December 3, 2002.
Since the wavelength calibration of XMM grating spectra  strongly depends
on the position of the 0th order, I used the VLBI position as centroid 
of the 0th order source \citep{ma98}. 
Extraction regions, for source and background, were chosen to be, 
respectively, within the 95\% and outside the 98\% of the
PSF. To exclude high particle background 
periods caused by solar activity, I extracted the 
background lightcurves from CCD-9 and 
excluded all the time intervals for which the background count rate
was higher than $1.5 \times 10^{-3}$ count s$^{-1}$. The net 
exposures for each observation are reported in Table~1.

\begin{figure}
\plotone{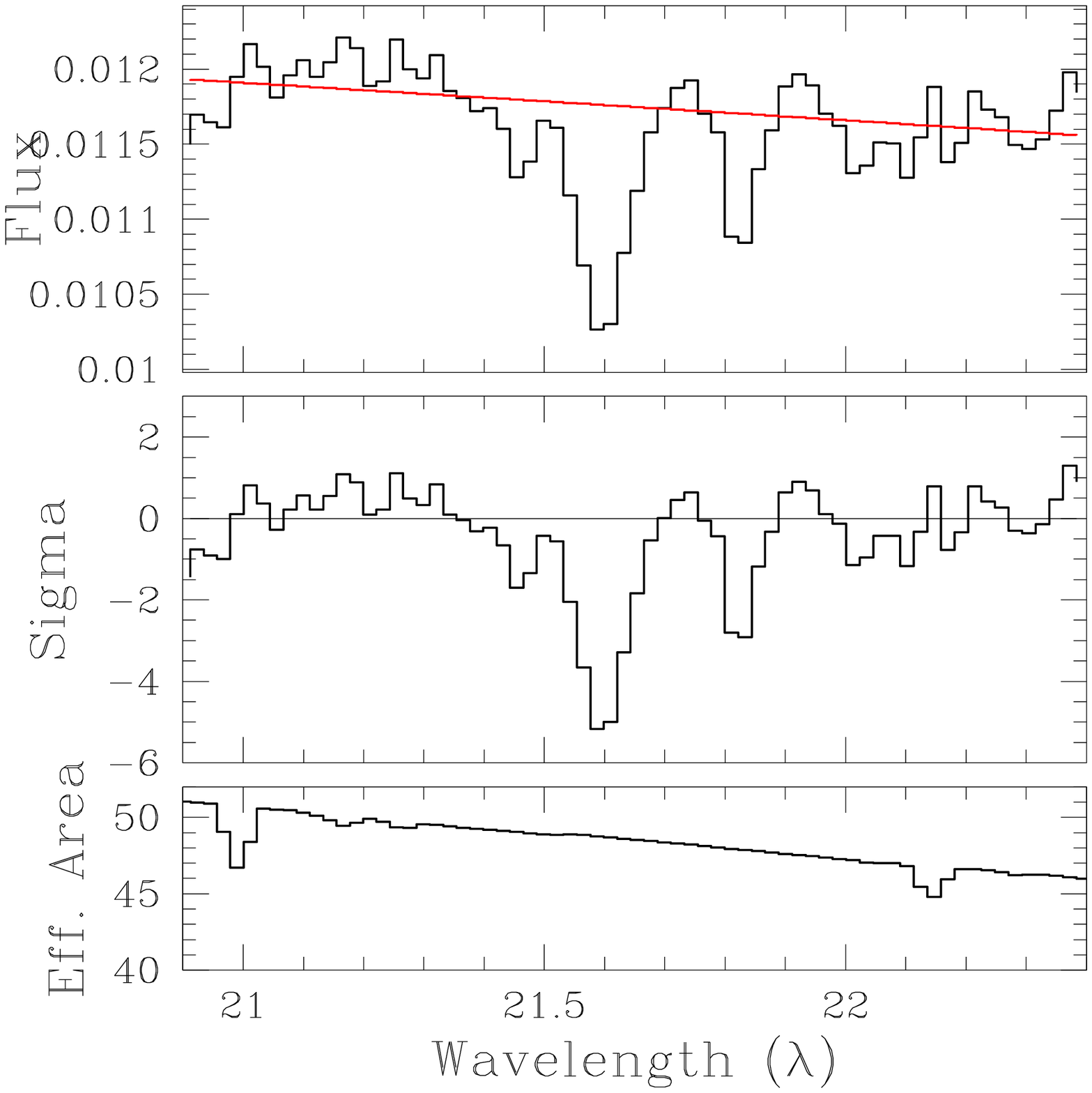}
\caption{Top panel: MRK~421 20.9--22.6 \AA \/ combined RGS-1 spectrum and best fit absorbed power law. Middle panel: residuals to the fit and, bottom panel, RGS-1 combined effective area.\label{fig1}}
\end{figure}

In order to improve the signal to noise ratio (SNR) I 
combined all the RGS-1 order 1 spectra obtained with a pointing offset $< 15^{\prime \prime}$
\footnote{The inclusion of the off-axis data in the combined spectrum
increases the number of photons at 20 \AA \/ from $\sim 4000$ to $\sim 7000$
per 0.06 \AA \/  resolution element, but degrades the
quality of the spectrum.  May 2002 observations, for example,
are highly affected by hot columns and pixels which result in
 spurious features in and around the expected  WHIM lines.
To reduce the impact of hot columns and pixels, the RGS units have recently 
been cooled (November 2002).}
(see Table~1)
and computed the  combined RGS-1 response matrixes 
(a convolution of the detector response matrix and effective area, 
Fig.~\ref{fig1}) using the PINTofALE interactive data language software
suite \citep{kas00}. 
I did the same for RGS-2.
The total net exposures are 125~ks and 122~ks for RGS-1 and RGS-2 respectively.

\section{The WHIM detection}\label{sec_whim}

\begin{figure}
\plotone{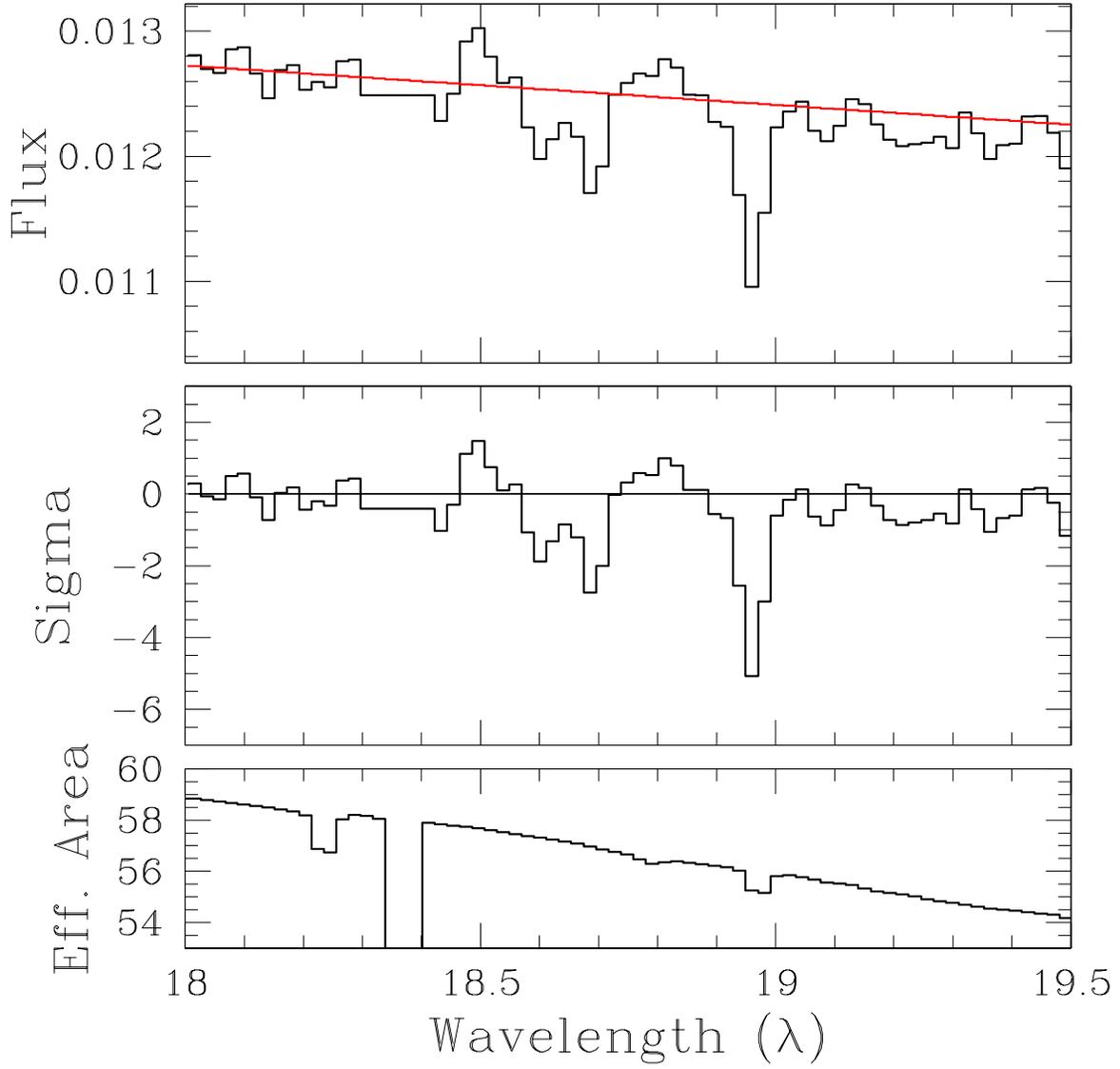}
\caption{Same as Fig.~\ref{fig1} for the 18.0--19.45 \AA \/ region.
\label{fig2}}
\end{figure}

I fit the combined fluxed RGS-1 spectrum
in the wavelength range 14.2--38.1 \AA \/ 
using version 2.2 of 
SHERPA (Siemiginowska et al., in prep.) modeling and fitting tool 
from the CXC analysis package CIAO 2.2 (Elvis et al., in prep.).
I used an absorbed power law model  with 
absorption characterized by  a column density N(X) and an
absorption cross section $\sigma_{\rm{X}}$ for each element.
I included in the model H, He {\sc i} and He {\sc ii} \citep{rum94}
and heavier elements \citep{mor83}.
I fixed the Galactic hydrogen column density at $N_H =1.61 \times
10^{20}$~cm$^{-2}$ \citep{loc95} and the ratios
$N_{He{\rm I}}$/$N_{H{\rm I}}$=0.1  and $N_{He{\rm II}}$/$N_{H{\rm I}}$=0.01. 
The best fit RGS-1 photon slope is $\Gamma _\lambda = 0.095 \pm 0.004$   
(corresponding to an energy photon slope of $2.095 \pm 0.004$) and the 
normalization is 0.0136  photons cm$^{-2}$ s$^{-1}$ \AA $^{-1}$ at 20.5 \AA .
In order to properly model the continuum around the expected WHIM features,
 I performed local absorbed power-law fits in the 
18.0--20.6 \AA \/ and in the 20.9--22.4 \AA \/ regions.
I kept the absorbing column fixed to the Galactic value in MRK~421 direction
and modeled the absorption lines with Gaussian.
The lines detected at  $>2 \sigma$ and the corresponding best fit parameters are listed in Table~2.\\
It appears to us that the only possible interpretation of 
  the  21.82 \AA \/ line ($2.8 \sigma$ detection; Fig.~1) 
is  redshifted  O~{\sc vii} K$\alpha$. This places the gas producing it at
$z \sim 0.01$ and makes this a solid X-ray detection of WHIM outside our local group of Galaxies.
The corresponding $z\sim 0.01$ O~{\sc vii} K$\beta$ is  expected to be too faint 
(EW$\sim 0.8$~m\AA \/ at $\sim 18.84$ \AA ) to be detected in this spectrum.
If any  O~{\sc viii} were present in the  $z \sim 0.01$ WHIM, the strongest expected 
line would be Ly$\alpha$ at 19.18 \AA , not detected in our spectrum.
The $3 \sigma$ upper limit on the line EW, computed fixing the line FWHM 
to the 21.82 \AA \/ line FWHM value,  is 2.47 m\AA . 
Under the assumption of unsaturated line the EW ratios between different 
species depend on the gas temperature and density.
Using  the ratio between the upper limit on the EW of  O~{\sc viii} at 
$z\sim 0.01$ and the 21.82 \AA \/ line EW, I obtain an upper limit on the 
$z\sim 0.01$ gas temperature of $\sim 4 \times 10^6$~K, for a  gas density 
of 1 atom cm$^{-3}$ and of  $\sim 2.5 \times 10^6$~K, for a  gas density 
of $10^{-6}$ atom cm$^{-3}$ \citep[see Fig.~5 in][]{nic02}.
These values are consistent with the temperature range predicted for the 
WHIM emitting in the soft X-ray band.
I derive from the 21.82 \AA \/ line a velocity of $cz \sim 3062$ km s$^{-1}$, 
consistent, within the errors, with the redshift of a strong H Ly$\alpha$ 
line detected by HST on MRK~421 spectrum ($cz = 3035 \pm 6$ km s$^{-1}$) and coming 
 from absorbing material in a cosmic void \citep{shu96,pen00}.
The location of WHIM in a cosmic void is consistent with the picture that 
such gas is connecting the overdense regions of the sky that collapsed 
into clusters and groups during the  structures formation.\\
For the other lines listed in Table~2, the  strongest  is the O~{\sc vii} 
K$\alpha$ absorption at  21.6 \AA \/ (Fig.~\ref{fig1}); the corresponding  
O~{\sc vii} K$\beta$ is visible at 18.67 \AA  \/ (Fig.~\ref{fig2}). 
The 21.6 \AA \/  feature has  already been seen in MRK~421 itself \citep{nic01} 
and in other 4 AGNs \citep{nic02,fan02,cag03,mat02,fan03} and it is usually attributed to 
a WHIM within our local group of galaxies \citep[e.g.][]{nic02}
or to radiatively cooling gas inside our Galaxy \citep[e.g.][]{hec02}.
I also report the  detection of zero redshift O~{\sc viii} Ly$\alpha$ at 
18.97 \AA \/ (Fig.~2) and of  Ne{\sc ix} K$\alpha$ at 13.4 \AA \/ (Fig.~3).
Note however that these lines are affected by small features in the effective area.\\
The EW of the zero redshift features (O~{\sc vii},O~{\sc viii} and Ne{\sc ix})
are compatible with those measured with {\em Chandra}  for PKS2155-304 
\citep{nic02,fan02} and for MRK~421 \citep{nic01}. In particular, 
the EW derived fixing the lines FWHM to the value of the 21.6 \AA \/ line, 
(i.e. 0.33, 0.27 and 0.20~eV respectively) 
are consistent with the  lines being unsaturated and with  column densities of 
$\sim 10^{16}$ atoms cm$^{-2}$ per ion species \citep[see Fig.~4 in][]{nic02}.
For a gas density of $\sim 10^{-6}$ atoms cm$^{-3}$, 
 Ne{\sc ix}/O{\sc vii} and O~{\sc viii}/O{\sc vii} ratios indicate gas 
temperatures of $\sim (0.4-2) \times 10^6$~K, and $\sim (1-4) 10^6$~K, respectively  \citep[Fig. 5 of][]{nic02}.
These ranges move to  $\sim (1.5-2.5) \times 10^6$~K and $\sim (3-4) \times 10^6 $~K 
in case of Galactic gas density.
Even if these estimates have to be regarded with caution because of the possible 
modifications of the effective area features, 
they  are all consistent with the range of temperatures predicted for the WHIM.

\section{Conclusion}\label{sec_sum}
I report  strong evidence of WHIM outside our local group of galaxies through
 the detection of a 21.82 \AA  \/ O~{\sc vii} K$\alpha$ absorption  
line in the   {\em XMM-Newton} RGS-1 spectrum of MRK~421.
The line redshift is $\sim 0.01$,  in agreement with that derived 
for the strong H Ly$\alpha$ absorption line detected by HST and related 
to an absorber in a cosmic void. The three closest galaxies are at 
2.14, 3.98 and 4.12 $h_{70}^{-1}$~Mpc \citep{pen00}. 
This detection is a firm proof the existence of a WHIM filament at $z\sim 0.01$
possibly connecting the nearby galaxies with other denser regions of the sky, 
as expected by the models of the cosmic structures formation and evolution.
I derive an upper limit on the gas temperature of $\sim 2.5-4 \times 10^6$~K 
for a gas density of $10^{-6} - 1$ atom cm$^{-3}$.
The only known similar evidences are  a  $4.5 \sigma$ detection of  
O~{\sc viii} at $z\sim 0.05$ reported by \citet{fan02} in the {\em Chandra} 
spectrum of PKS~2155-304 and the $\sim 2 \sigma$ detections of two O~{\sc vii} 
lines at $z\sim 0.2$ in the {\em Chandra} spectrum of H1821$+$643 \citep{mat02}.
However \citet{fan02} feature is not confirmed in other {\em Chandra}  and 
{\em XMM-Newton} observations \citep[e.g.][]{nic02,cag03}.\\
I also detect the zero redshift absorption features of  O~{\sc vii}, O{\sc vii} and Ne{\sc ix}  
with positions and EWs consistent with previous detections in the direction of other bright AGNs \citep[e.g.][]{nic02,cag03}.\\

\acknowledgments

I  acknowledge a C.N.A.A. fellowship.
I thank Aldo Treves and Francesco Haardt for useful scientific 
discussions and a careful reading of the manuscript.

\clearpage 

\clearpage 

\clearpage 

\begin{figure}
\plotone{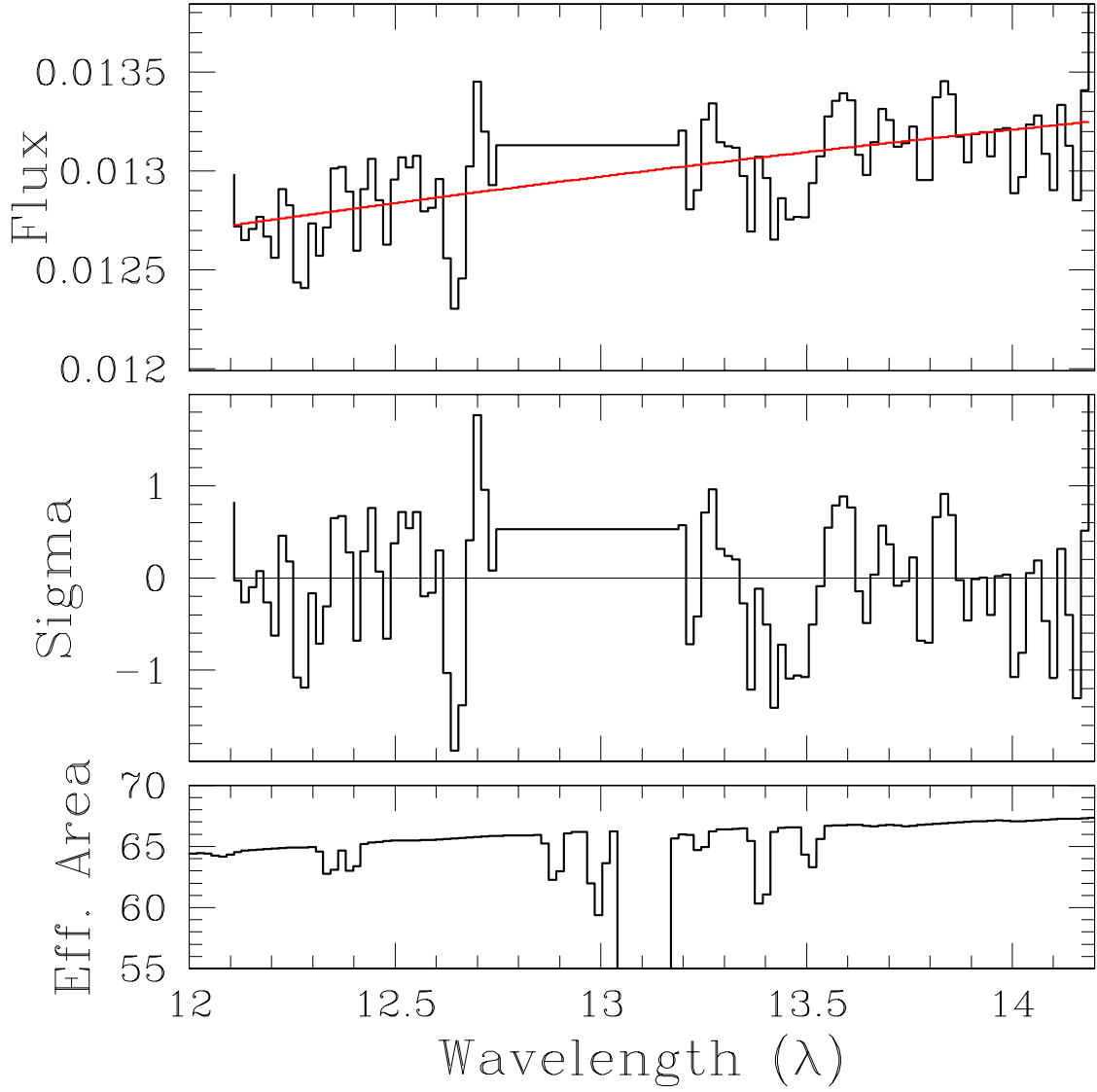}
\caption{Same as Fig.~\ref{fig1} for the RGS-2 vawelength range 12.0--14.2 \AA . \label{fig3}}
\end{figure}

\clearpage

\begin{deluxetable}{c ccc c c}
\tablecaption{All the public {\em XMM-Newton} RGS observations of MRK~421.}
\tablewidth{0pt}
\tablehead{
\colhead{Obs. Id.}
& \colhead{Obs. Start time\tablenotemark{a}}   &\colhead{Total Exposure\tablenotemark{b}} 
&\colhead{Net Exposure\tablenotemark{c}} &\colhead{Offset} &\colhead{Used observartion?\tablenotemark{d}}
}
\startdata
0099280201 &00-11-01 	&40 	&34  	&6.3	&YES\\
0099280301 &00-11-13 	&50 	&40 	&5.87	&YES\\
0099280401 &00-11-14 	&43 	&27 	&116.69	&NO\\
0099280501 &00-11-13    &21  	&15  	&7.63	&YES\\
0099280601 &00-11-15	&20	&12	&117.4	&NO\\
0136540101 &01-05-08	&39	&36	&12.95	&YES\\
0136540201 &01-05-08	&10	&0	&13.16	&NO\\
0153950601 &02-05-04	&40	&39	&123.98	&NO\\
0153950701 &02-05-05	&20	&18	&169.43	&NO\\
0153950801 &02-05-05	&22 	&21	&177.49	&NO\\
\enddata
\tablenotetext{a}{YY-MM-DD}
\tablenotetext{b}{Total RGS-1 ($\sim$RGS-2) on time in ks.}
\tablenotetext{c}{For RGS-1 in ks after the high particle background times rejection. 
RGS-2 net exposure times are similar.}
\tablenotetext{d}{For details see footnote~2 in the text.}
\end{deluxetable}

\clearpage

\begin{table}
\begin{center}
\caption{Best-fitting RGS-1 absorption line parameters and $1 \sigma$ errors.}
\begin{tabular}{cccccc}
\tableline\tableline
Line ID &$\lambda$ &$cz$ &FWHM &EW\tablenotemark{a} &$\sigma$\\
&(\AA ) &(km s$^{-1}$) &($10^{-2}$ \AA ) &(m\AA ) &\\
\tableline
Ne{\sc ix}\tablenotemark{b}
&$13.441 \pm 0.029$	&$-163^{+659}_{-616}$ 	&$12.06^{+4.71}_{-11.43}$ &$3.52^{+1.70}_{-1.47}$ ($2.89^{+1.30}_{-1.23}$) &2.5\\
O{\sc vii} K$\beta$
&$18.659^{+0.023}_{-0.019}$	     &$486^{+376}_{-299}$   &$11.22^{+2.80}_{-10.07}$ &$5.47^{+10.67}_{-1.63}$ ($4.88^{+1.29}_{-1.37}$)  &3.8\\
O{\sc viii} Ly$\alpha$\tablenotemark{b}
&$18.961 \pm 0.008$ 	     &$-98^{+41}_{-39}$  &$3.27^{+10.01}_{-17.34}$ &$5.50^{+5.72}_{-1.09}$ ($7.98^{+1.31}_{-1.33}$) &4.8\\
O{\sc vii} K$\alpha$	
&$21.597 \pm 0.008$ 	&$-75 \pm 80$ &$8.77^{+1.62}_{-1.33}$ &$12.67^{+1.59}_{-1.54}$ &7.6\\
\tableline
O{\sc vii} K$\alpha$
&$21.822^{+0.027}_{-0.024}$	 &$3062^{+333}_{-375}$ &$2.77^{+0.83}_{-1.25}$  &$3.68^{+4.70}_{-0.70}$ &2.8\\
\tableline
\end{tabular}
{\footnotesize
\tablenotetext{a}{The values in parenthesis are obtained fixing the FWHM of the line to the value of the zero redshift O{\sc vii} K$\alpha$}
\tablenotetext{b}{These lines could possibly be enhanced or modified  by  the presence of small effective area features (see the bottom panels in Fig.~2 and Fig.~3)}
}
\end{center}
\end{table}


\begin{thebibliography}{}
\bibitem[Cagnoni et al.(2003)]{cag03} Cagnoni, I. et al., 2003, ApJ in preparation
\bibitem[Fang \& Canizares(2000)]{fan00} Fang, T.~\& Canizares, C.~R.\ 2000, \apj, 539, 532 
\bibitem[Fang et al.(2002)]{fan02} Fang, T., Marshall, H.~L., 
Lee, J.~C., Davis, D.~S., \& Canizares, C.~R.\ 2002, \apjl, 572, L127 
\bibitem[Fang, Sembach \& Canizares(2003)]{fan03} Fang, T. ,  Sembach, K. R. \&  Canizares, C. R., 2003, ApJL in press astro-ph/0210666
\bibitem[Fukugita, Hogan, \& Peebles(1998)]{fuk98} Fukugita, M., Hogan, C.~J., \& Peebles, P.~J.~E.\ 1998, \apj, 503, 518 
\bibitem[Kashyap \& Drake(2000)]{kas00} Kashyap, V. \& Drake, J. J., 2000, Bull. Astron. Soc. India, 28, 475
\bibitem[Heckman et al.(2002)]{hec02} Heckman, T. M. ,  Norman, C. A., Strickland,  D. K. and Sembach,  K. R., 2002, ApJ in press (astro-ph/0205556)
\bibitem[Hellsten, Gnedin, \& Miralda-Escud{\' e}(1998)]{hel98} Hellsten, U., Gnedin, N.~Y., \& Miralda-Escud{\' e}, J.\ 1998, \apj, 509, 56 
\bibitem[Lockman \& Savage(1995)]{loc95} Lockman, F.~J.~\& Savage, B.~D.\ 1995, \apjs, 97, 1 
\bibitem[Ma et al.(1998)]{ma98} Ma, C.~et al.\ 1998, \aj, 116, 516
\bibitem[Mathur, Weimberg \& Chen(2002)]{mat02} Mathur, S., Weimberg, D. H. \& Chen, X., 2002,  proceedings of the conference "IGM/Galaxy Connection- The Distribution of     Baryons at z=0", astro-ph/0210575
\bibitem[Morrison \& McCammon(1983)]{mor83} Morrison, R.~\& McCammon, D.\ 1983, \apj, 270, 119 
\bibitem[Nicastro et al.(2001)]{nic01} Nicastro, F.~et al.\ 2001, (astro-ph/0102455)
\bibitem[Nicastro et al.(2002)]{nic02} Nicastro, F.~et al.\ 2002, \apj, 573, 157 
\bibitem[Nicastro, private comm(2002)]{nicpc} Nicastro, F., 2002, private communication
\bibitem[Penton, Stocke, \& Shull(2000)]{pen00} Penton, 
S.~V., Stocke, J.~T., \& Shull, J.~M.\ 2000, \apjs, 130, 121 
\bibitem[Rumph, Bowyer, \& Vennes(1994)]{rum94} Rumph, T., Bowyer, S., \& Vennes, S.\ 1994, \aj, 107, 2108 
\bibitem[Sembach et al.(2000)]{sem00} Sembach, K.~R.~et al.\ 
2000, \apjl, 538, L31 \bibitem[Sembay et al.(2002)]{sem02} Sembay, S., Edelson, R., 
Markowitz, A., Griffiths, R.~G., \& Turner, M.~J.~L.\ 2002, \apj, 574, 634 
\bibitem[Shull, Stocke, \& Penton(1996)]{shu96} Shull, J.~M., 
Stocke, J.~T., \& Penton, S.\ 1996, \aj, 111, 72 
\bibitem[Tripp et al.(2001)]{tri01} Tripp, T.~M., Giroux, 
M.~L., Stocke, J.~T., Tumlinson, J., \& Oegerle, W.~R.\ 2001, \apj, 563, 
724 



\end{thebibliography}
\end{document}